# Electrostatic Force Microscopy on Oriented Graphite Surfaces: Where Insulating and Conducting Behaviors Coexist

Yonghua Lu[1], M. Muñoz[1], C. S. Steplecaru[1], Cheng. Hao[1], Ming Bai[1], N. Garcia[1], K. Schindler[2], and P. Esquinazi[2]

[1] *Laboratorio de Física de Sistemas Pequeños y Nanotecnología, Consejo Superior de Investigaciones Científicas (CSIC), Serrano 144, 28006 Madrid, Spain*

[2] *Division of Superconductivity and Magnetism, University of Leipzig, Linnéstrasse 5, D-04103 Leipzig, Germany*

(Dated: April 10, 2006)

Abstract

We present measurements of the electric potential fluctuations on the surface of highly oriented pyrolytic graphite using electrostatic force and atomic force microscopy. Micrometric domain-like potential distributions are observed even when the sample is grounded. Such potential distributions are unexpected given the good metallic conductivity of graphite because the surface should be an equipotential. Our results indicate the coexistence of regions with metallic and insulating behaviors showing large potential fluctuations of the order of 0.25V. We discuss the implications of these measurements in the disorder structure of graphite.

The paradigm of anisotropic systems in nature, graphite, finds nowadays an extraordinary revival in the solid state physics community. The magnetic field driven metal-insulator-like transition in the basal resistivity of highly oriented pyrolytic graphite (HOPG) samples[1], followed by the discovery of the quantum Hall effect[2] (QHE) as well as evidence for Dirac fermions[3] (massless particles due to the linear dispersion relation) leave little doubt that the classical view of the physics of its magnetotransport properties is far from being adequate. These observations were confirmed on bulk graphite[4], a few layers thick graphite[5], and in graphene[6] (single graphite layer) where the contribution of the Dirac fermions makes the plateaus in the QHE to occur at half-integer filling factors. The occurrence of the QHE in bulk graphite suggests that the coupling between graphene layers should be much smaller than the one assumed by the Slonczewski-Weiss-McClure theory[7-9] but agrees with the results obtained by Haering and Wallace.[10] Nevertheless and in spite of a large number of experimental work on the transport properties of bulk graphite and graphene, their understanding is far from being satisfactory.

For example, different contact distributions on the surface of a HOPG sample show a small but striking influence on the measured properties. It appears that the internal disorder influences the field and temperature dependence of the transport properties in such an extent that, for example, the QHE can in some cases transform in a linear field dependent Hall effect with minima at specific fields.[2] In a recent experimental work a Corbino disk geometry was prepared on the surface of a high quality HOPG sample.[11] Using a voltage configuration where for a homogeneous sample no Hall signal is expected, clear evidence for an integer QHE was obtained. Based on the disordered semiconductor model of Ref. 12 an explanation for these results was proposed considering that the current paths are non-homogeneously distributed in the sample, in spite of its high quality. In agreement with this model the magnetoresistance at high enough fields appears to be governed by the Hall resistance,[11] providing a possible solution to the long-standing problem of the quasi-linear and non-saturating magnetoresistance in graphite.[13] In addition, recent theoretical work emphasizes the influence of edge states[14], curved surfaces and pentagonal or heptagonal defects[15] on the electronic states of graphene sheets. These "inhomogeneities" would imply the formation of a kind of domain-like structure. Also, electric screening is expected to be poor in graphene sheets[16] even if these are a "good conductor" at room temperature, assuming that a "good conductor" has low resistivity.

In this work we use the Electrostatic Force Microscopy (EFM) technique to characterize the electric potential distribution on the surface of a HOPG sample of the same high quality as in Ref. 11. Taking into account the low resistivity of the sample (basal resistivity at room temperature $\rho_b \leq 40$ $\mu\Omega$cm) and its clear metallic temperature dependence one would expect no voltage variation on the surface, or a continuous potential gradient in case the edges of the sample are connected to a potential difference and a current is passing through. In contrast to this expectation our results show huge voltage differences on the graphite surface, between regions of micrometer size in which the voltage appears to be rather constant. We propose that this surprising result - for a nominally conducting sample - may have its origin in the microstructure of the sample and the sensitivity of the electronic system of the graphite sheets to defects and/or deformation of the lattice. Our results reveal an unknown behavior of graphite and warn on interpretations of transport data based on assumptions of translation invariance and homogeneous current and voltage distributions, which may apply also for single graphene layers with disorder.

The sample measured in this work is a HOPG from Advanced Ceramics Co. with FWHM of the rocking curve width (mosaicity) of 0.4°.The anisotropy ratio (c-axis resistivity divided by the basal resistivity) is ~$10^4$ at room temperature. The total magnetic impurity concentration (measured with different methods) is less than 1ppm (Fe impurities less than 0.2 ppm). In our experiment, an AFM (NTI Solver) is used to measure the electrostatic force between a conductive probe and the surface of HOPG. The sample is freshly cleaved by simply peeling off the top layer using scotch-tape and electrically grounded on the top layer. Electrostatic force microscopy are performed in two-pass mode: the first pass (tapping mode) gives the topography of the surface, the second pass follows the track gotten in the first pass with a fixed lift-up distance and measures the frequency shift caused by the electrostatic force while the tip is biased by a voltage of a few volts. We used conductive, $W_2C$ coated tips. The tip is set to vibrate with a piezo having a resonant frequency of 150.09 kHz. This technique provides the surface topography and the potential fluctuations on the sample.

With the EFM technique one measures a shift of the vibration frequency of the cantilever with the conducting tip given by the interaction force between this and the sample. The force between scanning tip and a conducting material due to the applied bias voltage on the tip can be understood simply with a capacitor model. Supposed the capacitance of the tip-sample system is C, the attractive z-component of the force can be written as

$$F_z = \frac{1}{2}\frac{\partial C}{\partial z}(\Delta U)^2 = \frac{1}{2}\frac{\partial C}{\partial z}(U_{tip} - \varphi(x, y))^2 \quad , \tag{1}$$

where $U_{tip}$ is the voltage applied to the cantilever and $\varphi(x,y)$ is the potential distribution on the sample surface. If we apply a constant voltage to the tip and scan it at a constant distance from the sample, then the measured EFM signal indicates the potential fluctuation on the sample. In the experiment, the frequency shift from the resonance is measured, which depends linearly on the force gradient given by

$$\frac{\partial F_z}{\partial z} = \frac{1}{2}\frac{\partial^2 C}{\partial z^2}(U_{tip} - \varphi(x, y))^2 \quad . \tag{2}$$

In order to verify this capacitor mode, we measured the force gradient signal, indicated as "magsin" in the instrument (in nA units, linearly related to the frequency shift), with a tip-graphite surface distance of 100nm, while changing the biased voltage on the tip from -5V to 5V. Figure 1(a) shows the parabolic relation between "magsin" and bias voltage in two different regions of the sample surface, in agreement with Eq.(2). The proportionality between the square root of "magsin" and the bias voltage (see Fig. 1(b)), which changes for each working tip-surface distance, is used to obtain the potential distribution on the sample.

One expects that for a conducting sample with translational invariance the EFM signal should be constant on the sample surface. This is what we observe when we measure with the EFM a metallic gold surface. In fact this is true in a few, small areas of the graphite surface. However, to our surprise, most of the graphite surface shows a domain-like signal, see Figs. 2(c,d). To be more specific, the EFM measurement is repeated around 10 times on the same area with 3V voltage on the tip and 100nm lift-up distance. The scans show bright (insulating-like) spots forming regions with a size of several micrometers. As an example, Figs. 2(c) and (d) show the results of the $4^{th}$ and $5^{th}$ scans. The EFM images clearly show large potential fluctuations of about 0.25V, while the topography consists of flat terraces with a few nanometer curvature in the range of a few microns, see Figs. 2(a) and (b). This curvature is small but may be important for the electronic properties. The result shown in Fig. 2 is extremely interesting because we have large potential fluctuations, which are rather typical for an insulator with localized charges. On the other hand the sample we measure is a very good conductor and this should imply that the surface is an equipotential, as we have verified for a Au surface.

There are two factors that contribute to the EFM signal. One is the capacitance between the tip and sample, determined by the tip-sample distance, the material and shape of sample and tip. The other is the potential difference between the tip and sample. Generally, the voltage on the tip is constant, so the potential difference depends linearly on the sample potential. Reversing the polarization of the tip voltage we can discriminate between the influence of the capacitance or of the potential variation. If the signal is due to a capacitance variation, EFM images taken with positive or negative tip bias voltage should be the same. Otherwise, the EFM images with positive or negative bias voltage should be complementary. Assume that there are two regions on the sample surface with potentials $V_1$ and $V_2$ ($V_1 > V_2$). If we apply a positive voltage V to the tip, the potential difference on the regions 1 and 2 is $(V-V_1) < (V-V_2)$. If the tip voltage is changed to negative (-V), then the potential difference is $(V+V_1) > (V+V_2)$. This means that the contrast of the EFM image will reverse when the polarity of tip voltage changes, which is the case in our experiment, see Fig.3. Therefore we state that the contrast of our EFM images is an indication of the potential distribution of graphite. The potential difference is estimated according to the linear relation shown in Fig.1(b).

To be more confident, we passed an electric current through the sample as well as changed the tip-sample distance to see what happen to the EFM images. Figure 4 shows the topography (a) and the EFM images taken when a current of 80 mA (b), then 100mA (c) and at last –100mA is applied to the sample. The potential fluctuations shown in Fig. 4(e) was calculated with the linear relation of Fig. 1(b) and obtained from the line scans done at the position of the black lines in Figs. 4(b) and (c). We note that the current has an interesting influence on the potential fluctuations but the main pattern appears stable. We have also performed EFM scans at two different lift-up heights, first at 100nm then 200nm. Figure 5(a) shows the topography and Figs. 5(b,c) the EFM images at the two heights. The potential difference is calculated taken into account the proportionality factors for the two heights shown in Fig. 5(d). Figure 5(e) demonstrates that the measured potential difference between the bright and black areas is practically the same for the two heights. Thus, the potential distribution on the surface of graphite is not induced temporarily by the biased tip while scanning. The observed potential fluctuation of 0.25 V is of the order of the assumed coupling between graphene layers used in Refs. 7-9 to explain the conductivity behavior of low anisotropic graphite samples. The low anisotropy is due to a higher defect density, which produces, in average, a better coupling between graphene layers in those samples.

Taking into account the experimental evidence discussed above, there are apparently contradictory physical results that puzzle the understanding of this amazing material. These are: i) it has a good conductivity but, theoretically, we expect that perfect graphene sheets should be low conducting (or not conducting at all at 0K), ii) there is the observation of the QHE that implies that HOPG samples show a bidimensional character, iii) on the other hand, one expects theoretically that a graphene sheet should be conducting due to defects. Our experimental results shed light on these apparent discordances in the following way. The graphite sample shows a domain-like structure with 0.25V variations in potential. We propose that these large potential differences are mainly due to the existence of two kinds of regions in graphite. Due to disorder and defects (not only at the surface) there are regions with conducting graphene layers that are well linked and coupled in the c-axis direction, providing good conducting regions that percolate. The other regions are of highly perfect graphene layers with small coupling between them (bidimensional character) that may be insulating-like and extend in a micron region. These last regions added to the domain structure provide in average the conditions for the observation of the QHE. Then the two contradicting points (i) and (ii) coexist. Finally, we note that the AFM pictures show a long range curvature that may imply the existence of defects as pentagons, heptagons or any others that can change the density of states and make particular regions conducting. This will provide an excess of charge on the curved graphene layers of the order of $10^{-4}$ electrons per atom changing their potential. Smaller fluctuations of the potential, also in bidimensional sheets, can be attributed to the influence of those defects and topology. The existence of defects will result also in the coexistence of $sp^2$ and $sp^3$ bonded carbon atoms that providing unsaturated spins that may couple producing a weak ferromagnetism as revealed by local probes magnetostriction measurements.[17]

In conclusion, EFM measurements on highly oriented pyrolytic graphite show potential fluctuations characteristic of an insulator and consistent with no or low electrical screening. On the other hand, the material is a good conductor. The way to explain these two behaviors is by the coexistence of two phases, one with defective graphene layers well coupled and regions with ideal layers with small coupling, as the EFM images indicate. The results help to understand the observation of the QHE in HOPG. In this picture also defects have to enter although it may be in a subtle way to provide a small curvature on the graphite terraces, as observed. We hope that these observations may well contribute to explain some puzzles of the transport properties of graphite as well as encourage researchers the use of the EFM technique for further characterization of this amazing material.

**Acknowledgments**
This work is supported by the European Union Ferrocarbon and BMR projects. One of us (K.S.) is supported by the DFG under grant ES 86/11-1. We thank M.A.Vozmediano, F. Guinea and M. Ziese for a carefully reading of the manuscript and for discussions.

Figure Captions:

Fig.1. (a) Parabolic dependence of the force gradient signal (called "magsin") on the tip bias voltage (BV) measured in two different regions of the sample. Dark and bright areas mean areas with high and low conductivity. The HOPG sample is electrically grounded. (b) After taking the square root of "magsin" a linear relation between this and the bias voltage can be used to convert the EFM signal into a potential distribution.

Fig.2. (Colour online) Electrostatic force microscopy measurements performed repeatedly on the graphite surface with +3V voltage on the tip. All images in this and further figures are taken in a 8 x 8 µm$^2$ area. (a) The topography measured with the AFM mode, (b) topography profile on the line indicated in 2(a). Notice the step of a few nm height as well as the small curvature of 3nm in ~ 7 µm length, (c) and (d) are the fourth and fifth EFM scanning.

Fig. 3. (Colour online) EFM images obtained after changing the polarity of the voltage applied on the tip. With tip voltage (a) BV = 3V and (b) with –3V. The obtained image indicates that the observed response is not due to a change in capacitance.

Fig. 4. (Colour online) Topography (a) and EFM images (b-d) obtained when an electric current through the sample of (b) 80mA, (c) 100mA and (d) –100mA was passing. (e) Line scans of the surface potential along the black lines shown in (b, c, d). We see that the main structure of the EFM signal is not influenced by the applied current in first approximation. Note that according to the sample resistivity the potential gradient in our scanning area is much less than 1mV, even when we use 100mA current.

Fig. 5. (Colour online) (a) Topography and EFM measurements done continuously at two different lift-up heights: (b) $z_1$ = 100nm and (c) $z_2$ = 200nm. (d) Linear relationship between the square root of the EFM signal and the on tip-sample voltage. (e) Line scans of the potential profile along the lines indicated in (b, c).

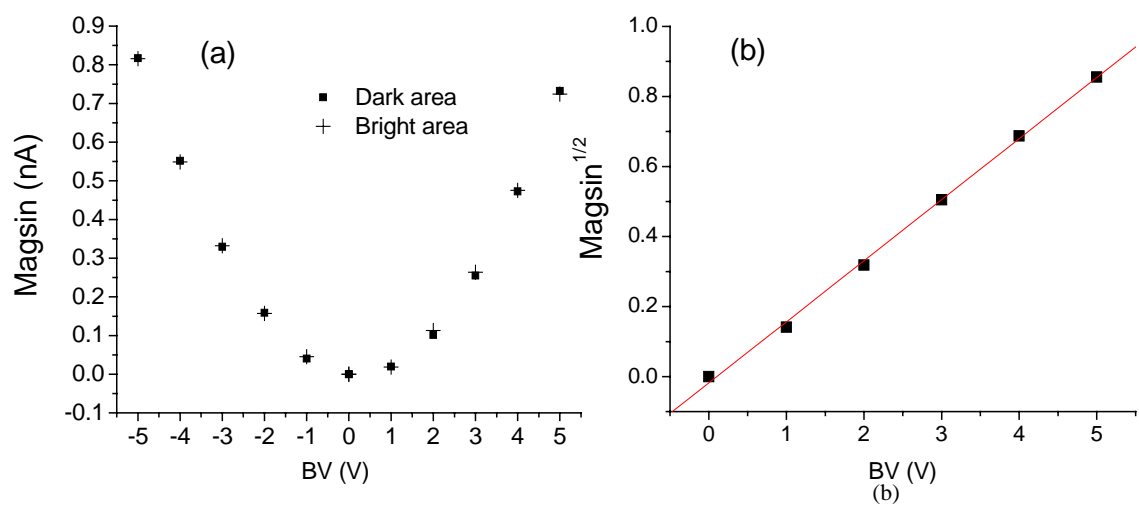

Fig. 1

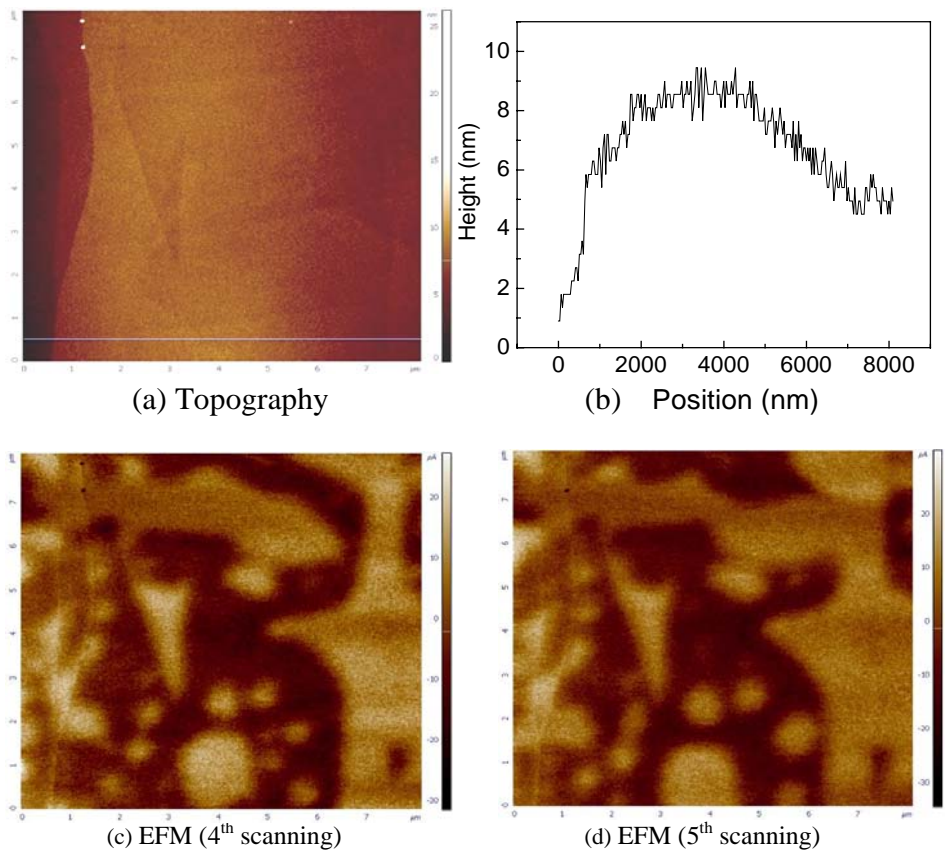

(a) Topography  (b) Position (nm)

(c) EFM (4[th] scanning)  (d) EFM (5[th] scanning)

Fig. 2

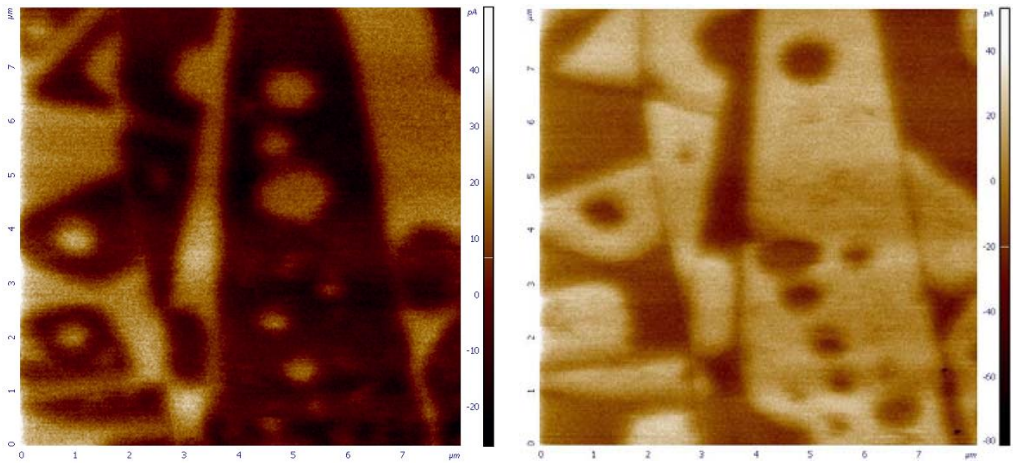

(a) BV=3V    (b) BV=-3V

Fig. 3

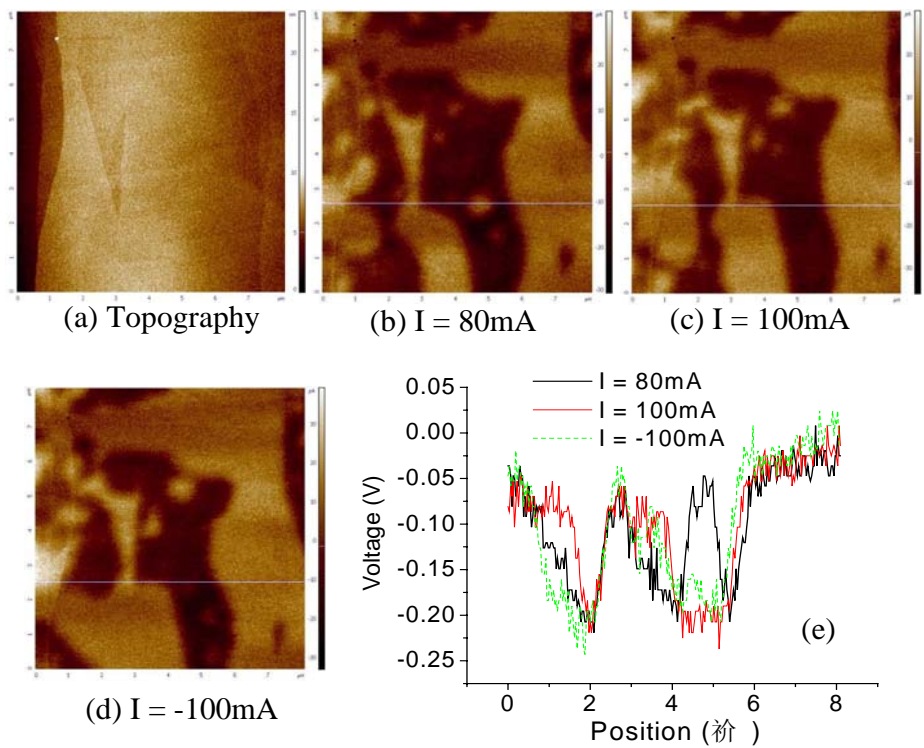

Fig. 4

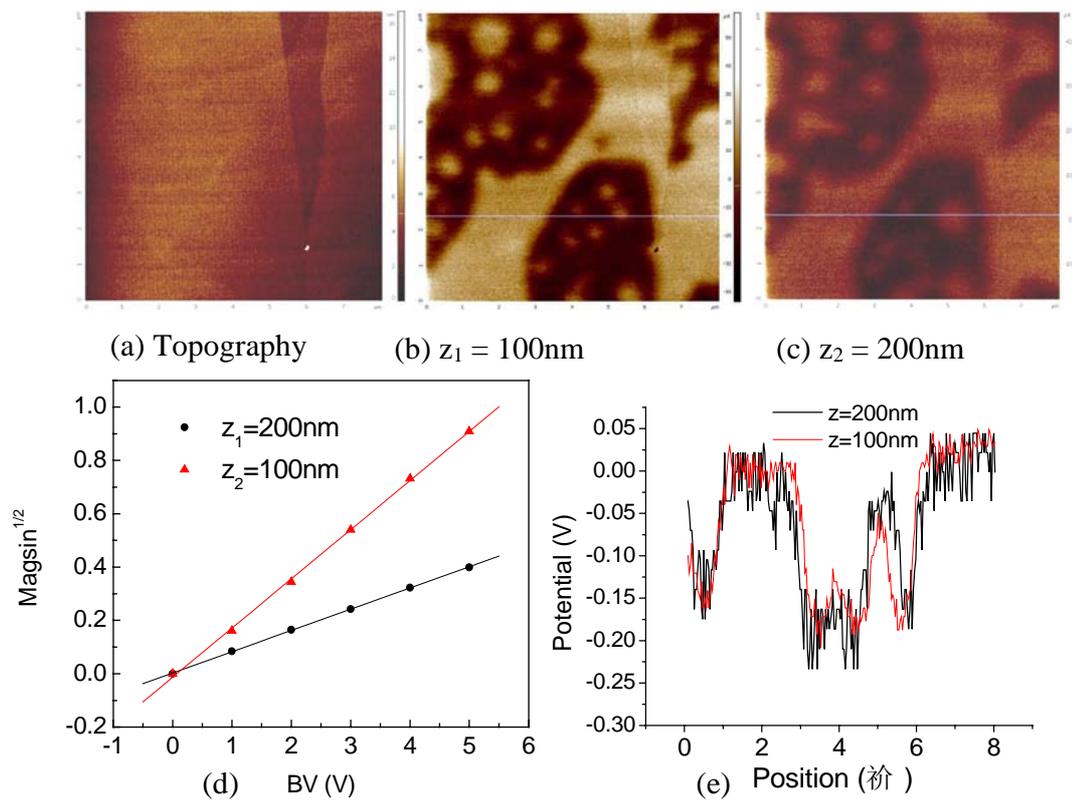

Fig. 5